\documentclass[doublecol]{epl2}
\usepackage{qsymbols}
\usepackage{amsmath}

\title{Experimental evidence of non-Amontons behaviour at a multi\-contact interface}
\shorttitle{Experimental evidence of non-Amontons behaviour at a multi\-contact interface}

\author{J. Scheibert\inst{1} \and A. Prevost\inst{1} \and J. Frelat\inst{2} \and P. Rey\inst{3} and G. Debr\'egeas\inst{1}}
\shortauthor{J. Scheibert\etal}

\institute{                    
  \inst{1} Laboratoire de Physique Statistique de l'ENS, UMR 8550, CNRS/ENS/Universit\'e Paris 6/Universit\'e Paris 7, 24 rue Lhomond, 75231 Paris, France\\
  \inst{2} Laboratoire de Mod\'elisation en M\'ecanique, UMR 7607, CNRS/Universit\'e Paris 6, 4 place Jussieu, 75252 Paris, France\\
  \inst{3} CEA-LETI, 17 rue des Martyrs, F38054 Grenoble Cedex 09, France
}


\abstract{We report on normal stress field measurements at the multicontact interface between a rough elastomeric film and a smooth glass sphere under normal load, using an original MEMS-based stress sensing device. These measurements are compared to Finite Elements Method calculations with boundary conditions obeying locally Amontons' rigid-plastic-like friction law with a uniform friction coefficient. In dry contact conditions, significant deviations are observed which decrease with increasing load. In lubricated conditions, the measured profile recovers almost perfectly the predicted profile. These results are interpreted as a consequence of the finite compliance of the multicontact interface, a mechanism which is not taken into account in Amontons' law.}

\begin{document}

\maketitle

Knowledge of the stress field at the contact region between two solids is of considerable interest to numerous domains such as mechanical engineering \cite{Hertz-1896, Johnson-BritJApplPhys-1958, Meijers-ApplSciRes-1968, Hardy-Baronet-Tordion-IntJNumMethodsEng-1971, Spence-JElasticity-1975, Johnson-CUP-1985}, solid friction \cite{Cattaneo-RANL-1938, Mindlin-JAppMech-1949, Hamilton-MechEngSci-1983} or seismology \cite{Scholtz-CUP-1990}. From a continuum mechanics point of view, the theoretical or numerical determination of this field requires a set of constitutive equations characterizing the mechanical response of the interface. These are usually inferred from macroscopic measurements: for instance, frictional contacts are often found to obey the empirical Amontons' friction law which states that irreversible sliding occurs when the ratio of tangential to normal forces reaches a static friction coefficient $\mu_{macro}$ without any prior deformation of the interface \cite{Baumberger-Caroli-AdvPhys-2006, Drees-Achanta-IJTC-2007}. The classical approach for calculating frictional contact stress field therefore consists in considering a smooth interface exhibiting an analog rigid-plastic response: $\mu_{macro}$ now defines the threshold ratio between shear and normal stress components for local slip to occur  \cite{Cattaneo-RANL-1938, Mindlin-JAppMech-1949, Spence-JElasticity-1975, Hamilton-MechEngSci-1983}.

Considering both a smooth interface and Amontons' law may seem paradoxical since the latter is expected to be valid only for rough solids. Because of adhesion forces, contact between molecularly smooth surfaces generally do not obey Amontons' friction law \cite{Vorvolakos-Chaudhury-Langmuir-2003, Bureau-Baumberger-Caroli-EurPhysJE-2006}. Furthermore, random roughness provides a microscopic basis to Amontons' law  \cite{Bowden-Tabor-OUP-1950, Greenwood-Williamson-ProcRSocLondA-1966, Brown-Scholz-JGR-1985, Persson-JChemPhys-2001}: it allows for the description of the interface as a collection of isolated load-bearing points forming a multicontact interface (MCI). For standard roughness characteristics, the real area of contact - and thus the tangential force required to trigger sliding - grows linearly with the applied load. 

When associated with Amontons' law, the smooth interface hypothesis must therefore be understood as the limiting case of a macroscopic surface bearing vanishingly small roughness. One may wonder to what extent this approximation is valid in real MCI's. Experimentally, this question has been addressed by focusing on the relationship between the macroscopic normal load and the indentation depth\cite{Nuri-Wear-1974, Kagami-Yamada-Hatazawa-Wear-1983, Brown-Scholz-JGR-1985, Benz-Rosenberg-Kramer-Israelachvili-JPhysChemB-2006}. Here we go further and compare the stress field measured within an extended MCI with Finite Elements Method (FEM) calculations assuming both smooth surfaces and the local (rigid-plastic) Amontons' law.

\begin{figure}
\includegraphics[width=\columnwidth]{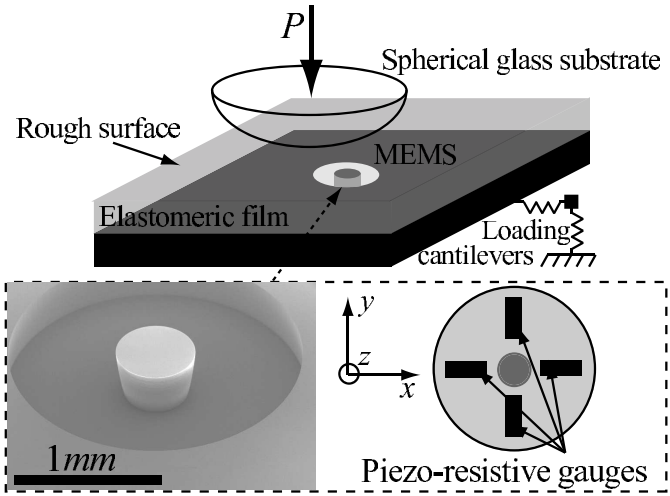}
\caption{Schematics of the stress sensor. A MEMS force sensor enables to measure the applied normal stress at the base of a rough, nominally flat PDMS film (thickness $h=2 \, mm$, lateral dimensions $50 \times 50 \, mm$) pressed against a spherical glass substrate. The macroscopic normal and tangential loads are measured through the extension of two orthogonal loading cantilevers (normal stiffness $641 \pm 5 \, N.m^{-1}$, tangential stiffness $51100 \pm 700 \, N.m^{-1}$) by capacitive position sensors (respectively MCC30 and MCC5, Fogale nanotech).\label{schemamanip-EPL1}}
\end{figure}

In order to perform such local measurements, we take advantage of the recent development of Micro Electro Mechanical Systems (MEMS) \cite{Kane-Cutkosky-Kovacs-SensorsActuatorsA-1996, Leineweber-Pelz-Schmidt-Kappert-Zimmer-SensorsActuatorsA-2000}. Local normal stress measurements are obtainend using a MEMS force sensor\footnote{This MEMS also measures the tangential stress, but with an accuracy which is not sufficient for the present study \cite{Scheibert-PhD-2007}.} embedded at the rigid base of a rough, nominally flat elastomeric film pressed against a rigid sphere under normal load (Fig.\ref{schemamanip-EPL1}). The MEMS sensitive part is a rigid cylinder (diameter $550 \, \mu m$, length $475 \, \mu m$) attached to a suspended circular Silicon membrane (radius $1 \, mm$, thickness $100 \, \mu m$) whose deformations are measured with four couples of piezo-resistive gauges (see inset of fig.\ref{schemamanip-EPL1}). The applied normal stress is therefore averaged over a surface of a few millimeters square.

The elastomeric material is a cross-linked poly\-di\-methyl\-siloxane (PDMS) (Sylgard 184, Dow Corning) of Young's modulus $2.2 \pm 0.1 \, MPa$, and of Poisson ratio $0.5$ \cite{PolyDataHandbook-OUP-1999}. No measurable stress relaxation being observed after a sudden loading, the PDMS can be considered as purely elastic. The film is obtained by pouring the cross-linker/PDMS melt on the MEMS into a parallelepipedic mold covered with a Poly\-Methyl\-Meth\-Acrylate plate roughened by abrasion with an aqueous solution of Silicon Carbide powder (mean diameter of the grains $37 \, \mu m$). After curing and demoulding, the resulting $rms$ surface roughness is measured with an interferential profilometer (M3D, Fogale Nanotech) to be $\rho=1.82 \pm 0.10 \, \mu m$. This roughness is sufficient to avoid any measurable pull-off force against smooth glass substrates, as discussed in \cite{Fuller-Tabor-ProcRSocLondA-1975}.

\begin{figure}
\includegraphics[width=\columnwidth]{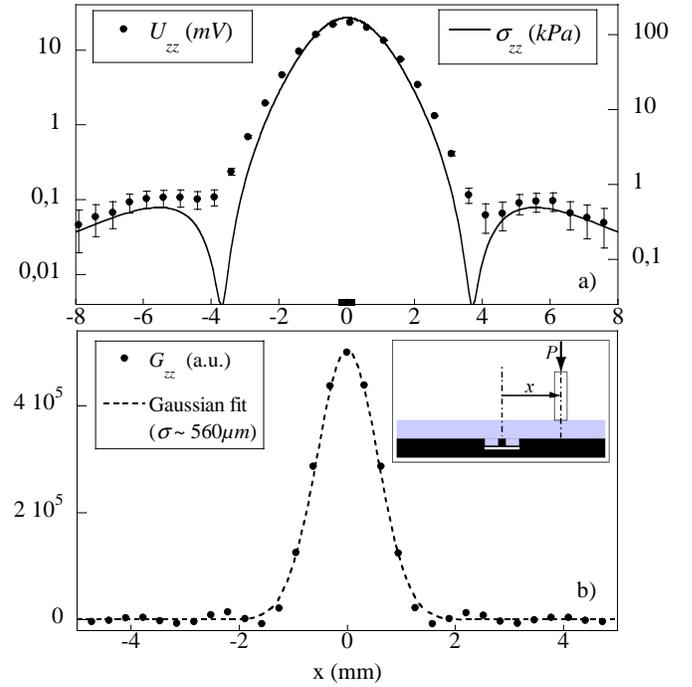}
\caption{a) Radial profiles associated to the indentation of the elastic film by a rigid thin rod, under normal load $P=1 \, N$. The measured output voltage $U_{zz}$ ($`(!)$) is compared to FEM results (solid line) for the normal stress $\sigma_{zz}$ at the base of the elastic film. The error bars represent the electronic noise. The black rectangular patch represents the rod diameter ($500 \, \mu m$). b) Normalized apparatus function of the MEMS for the normal stress, $G_{zz}$ ($`(!)$), as determined by numerical integration of eq.\ref{convol}, and its Gaussian fit in dashed line.\label{Calib-EPL1}}
\end{figure}

The stress sensing device is calibrated by indenting the film surface with a rigid rod of diameter $500 \, \mu m$, under a normal load $P$. With this type of indentor, the sensor output is found to be linear with the applied load. By successively varying the position of the rod along the $x$ direction, and assuming homogeneity of the surface properties of the film, one can construct point by point the normal output voltage radial profile $U_{zz}$ (fig.\ref{Calib-EPL1}a)). The latter is then compared to the results of Finite Elements Method (FEM, Software Castem 2007) calculations for the stress $\sigma_{zz}$ at the base of a smooth axi-symmetrical elastic film (of same elastic moduli and thickness) perfectly adhering to its rigid base and submitted to a prescribed normal displacement over a central circular area\footnote{Such results could have been obtained semi-analytically for frictionless conditions by using the model developed in \cite{Fretigny-Chateauminois-JPhysD-2007} but FEM calculations have been preferred here because they allow for variable boundary conditions.} of diameter $500 \, \mu m$. As expected for contact regions of dimensions smaller than the film thickness, the stress calculated at the base of the film is found to be insensitive to the frictional boundary conditions. 

Ignoring the stress field modifications induced by the MEMS 3D structure, one can relate the measured output voltage $U_{zz}$ to the stress field $\sigma_{zz}$ by writing down that
\begin{equation}
U_{zz}(x,y)=A_{zz} G_{zz} \otimes \sigma_{zz} (x,y) \label{convol}
\end{equation}
where $A_{zz}$ is a conversion constant (expressed in $mV / Pa$), $G_{zz}$ is a normalized apparatus function and $\otimes$ is a convolution product. In Fourier space, eq.\ref{convol} becomes
\begin{equation}
A_{zz} G_{zz}(x,y)=\mathcal{F}^{-1}\left(\frac{\mathcal{F}\left\{U_{zz}\right\}(f_{x},f_{y})}{\mathcal{F}\left\{\sigma_{zz}\right\}(f_{x},f_{y})}\right)(x,y) \label{Gzz}
\end{equation}
where $\mathcal{F}$ is the bidimensional spatial Fourier Transform, $\mathcal{F}^{-1}$ its inverse, and where $f_{x}$, $f_{y}$ are respectively the spatial frequencies in the $x$, $y$ directions. The $U_{zz}(x,y)$ and $\sigma_{zz}(x,y)$ fields are built from the profiles along the $x$ axis, assuming axisymmetry, and then transformed using the Fast Fourier Transform (FFT) algorithm. The fast decaying of $\mathcal{F}\left\{\sigma_{zz}\right\}$ with increasing spatial frequency introduces divergences of the ratio in eq.\ref{Gzz}. To avoid it, a white noise of amplitude ten times weaker than the weakest relevant spectral component is added to both terms of the ratio before applying the FFT. The result is found to be insensitive to the particular amplitude of this white noise. $A_{zz}$ is determined so that the integral of $G_{zz}$ is 1.

Figure \ref{Calib-EPL1}b) shows the resulting apparatus function. It has a bell shape with typical width of the order of $600 \, \mu m$ comparable to the MEMS lateral dimension. For the subsequent convolutions an approximated apparatus function, taken as a gaussian of standard deviation $561 \, \mu m$ (fig.\ref{Calib-EPL1}b)), has been used. This approximated apparatus function gives back the measured profile when convoluted by the calculated one, proving that neither the deconvolution method nor the approximation introduces significant loss of information. In particular, the point-like indentation involves a large enough spatial spectrum to allow for a faithfull determination of $G_{zz}$ within all of its relevant spatial components.

Sphere-on-plane MCIs are formed against an optical plano-convex spherical glass lens (radius of curvature $128.8 \, mm$). Both the glass and PDMS surfaces are passivated using a vapor-phase silanization procedure which lowers and homogenizes the surface energy. Each contact is obtained using the following loading sequence. The glass lens is pressed against the film up to the prescribed load $P$ within $2 \, \%$ relative error. Due to this loading, the extremity of the normal cantilever is tangentially displaced and a significant tangential load $Q$ is induced. Consequently, from this position, the contact is renewed by manual separation which results in a much smaller but finite $Q$. The glass lens is eventually translated a few micrometers tangentially down to $Q=0$. Both the surface treatment and the loading sequence are found to yield an excellent reproducibility of the measurements. As for the rod indentation, the radial profiles are derived from a series of $33$ contacts whose centers lie every $0.5 \, mm$ along the $x$ direction. These profiles divided by $A_{zz}$ have the dimension of a stress and are labelled $S_{zz}$.

For a quantitative comparison, FEM calculations are carried out for a frictional sphere-on-plane contact with the same geometry. Both contacting surfaces are taken as smooth and the interface is assumed to obey locally Amontons' friction law with a friction coefficient $\mu$. Both solids are discretized with a uniform mesh size of $50 \, \mu m$ and the normal displacement of the rigid elastic sphere is prescribed. The contact conditions are satisfied using a double Lagrange multiplier implying that both surfaces are slave and master. The normal load is reached step-by-step and at each step an iterative Newton-Raphson method is used to satisfy both the unilateral contact and the friction law.

\begin{figure}
\includegraphics[width=\columnwidth]{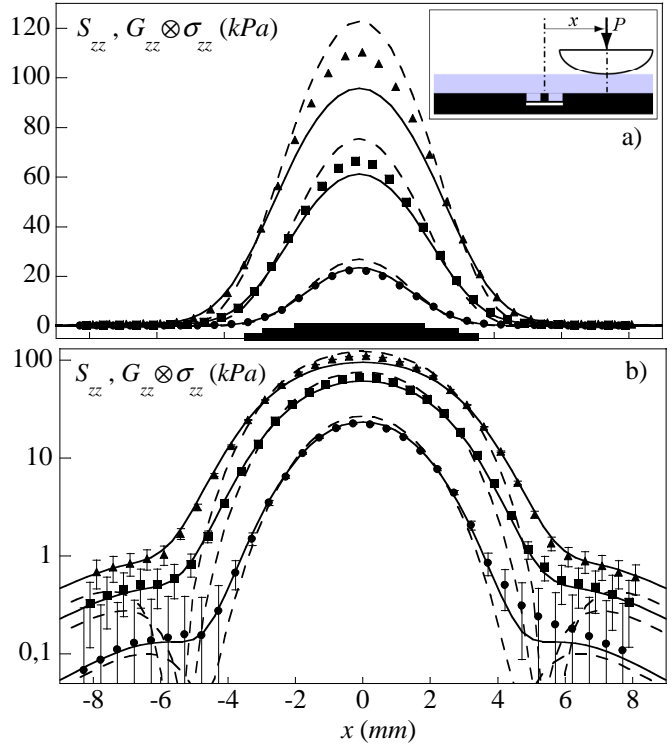}
\caption{Measured normal stress profiles ($S_{zz}$) under normal loading by a rigid sphere ($P=0.34 \, N$ ($\bullet$), $1.37 \, N$ ($`[!]$), $2.75 \, N$ ($\blacktriangle$)). Comparison is made with $G_{zz} \otimes \sigma_{zz}$ for $\mu=0$ (solid lines, indentation depths $18$, $33$ and $45 \, \mu m$ respectively) and $\mu=\infty$ (dashed lines, indentation depths $16$, $28$ and $37 \, \mu m$ respectively). a) Linear scale. b) Semi-logarithmic scale. The black rectangular patches on a) represent the contact diameters ($2.00$, $2.90$ and $3.45 \, mm$ for $P=0.34$, $1.37$ and $2.75 \, N$ respectively) obtained from the FEM calculations for $\mu=0$.\label{FIG-EPL1}}
\end{figure}

\begin{figure}
\includegraphics[width=\columnwidth]{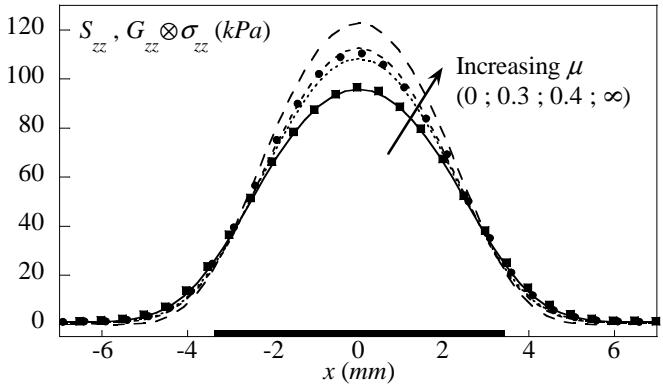}
\caption{Measured normal stress profile ($S_{zz}$) under normal loading ($P=2.75 \, N$) by a rigid sphere for both dry ($`(!)$) and glycerol-lubricated ($`[!]$) contacts. Shown in solid and dashed lines are the $G_{zz} \otimes \sigma_{zz}$ profiles for 4 values of $\mu$ ($0$, $0.3$, $0.4$, $\infty$). The black rectangular patch represents the contact diameter ($3.45 \, mm$) obtained from the FEM calculation for $\mu=0$ (indentation depth $45 \, \mu m$). \label{P2-EPL1-toutesf}}
\end{figure}

Figure \ref{FIG-EPL1} compares the $S_{zz}$ and $G_{zz} \otimes \sigma_{zz}$ profiles for two limiting boundary conditions, $\mu=0$ and $\mu=\infty$, and for three values of $P$. Within the error bar the measured profiles are bracketed by the two limiting numerical profiles over the whole spatial range and over 3 orders of magnitude, as clearly displayed on fig.\ref{FIG-EPL1}b). In the contact outer region, $S_{zz}$ is systematically very close to the frictionless profile whereas at the center it increasingly departs from it with the load, as discussed further. Similar measurements are performed, for two limiting loads ($P=0.69 \, N$ and $P=2.75 \, N$), under lubricated conditions. A glycerol droplet is inserted at the interface prior to loading. The profiles display both a larger spatial extent and a lower maximum amplitude than the ones obtained with dry conditions. They are correctly captured by the FEM calculation using a null friction coefficient (as shown for $P=2.75 \, N$ on fig.\ref{P2-EPL1-toutesf}).

\begin{figure}
\includegraphics[width=\columnwidth]{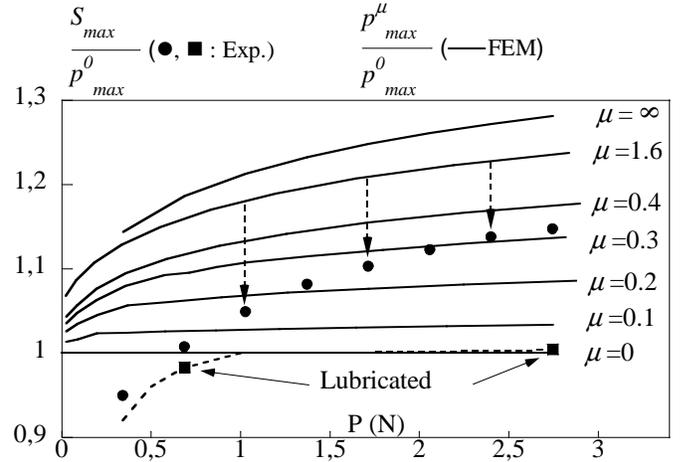}
\caption{Load dependence of $S_{max}$, the maximum normal stress measured for dry ($`(!)$) and lubricated ($`[!]$, guideline in dashed line) conditions, and $p^{\mu}_{max}$, the maximum calculated normal stress for various values of $\mu$. Both quantities have been normalized by $p^0_{max}$, the value of $p^{\mu}_{max}$ for $\mu = 0$. The vertical arrows represent the apparent drop of the friction coefficient from $\mu_{macro} ``~ 1.6$ down to $\mu_e$ for dry conditions.\label{Pmax-EPL1}}
\end{figure}

Except for the smallest load, any pressure profile under dry conditions can be correctly approached using an effective friction coefficient $\mu_e$ which can be determined by trial and error with a resolution of order 0.1 (as illustrated in fig.\ref{P2-EPL1-toutesf}). The load dependence of $\mu_e$ is exhibited by plotting the maximum pressure $S_{max}$ normalized by the maximum pressure $p^0_{max}$ calculated for a frictionless contact, for different loads. On the same graph, we plot the equivalent expression $p^{\mu}_{max}/p^{0}_{max}$ obtained by FEM calculation for different friction coefficients $\mu$ in the range $[0,\infty ]$. This representation allows one to directly \textit{read} the value of the effective friction coefficient $\mu_e$ for any load. $\mu_e$ is found to increase with $P$ but remains well below the macroscopic friction coefficient $\mu_{macro} \approx 1.6$ over the range of loads explored\footnote {the macroscopic friction coefficient has been measured for a driving velocity $v=100\mu m/s$ of the rigid base of the sensor: it shows a (small) decrease from $1.8$ to $1.5$ as the load is increased from $0.34 \, N$ to $2.75 \, N$, a behavior usually attributed to the finite adhesion energy of the interface \cite{Carbone-Mangialardi-JMechPhysSolids-2004}}.

The significant discrepancy between the effective and the macroscopic friction coefficient suggests that the rigid-plastic response of the interface underlying Amontons' law needs to be refined. One may in particular question the rigid hypothesis since MCI's are known to possess finite compliances both in normal and in-plane directions. The effect of the normal compressibility of a rough interface \cite{Greenwood-Williamson-ProcRSocLondA-1966, Brown-Scholz-JGR-1985, Benz-Rosenberg-Kramer-Israelachvili-JPhysChemB-2006, Persson-PRL-2007} on a sphere-on-plane contact has first been described by Greenwood and Tripp \cite{Greenwood-Tripp-JAppMech-1967}. They predict an increase of the apparent contact radius with respect to a smooth interface, as well as a decrease in the maximum normal stress. These deviations are expected to vanish when the ratio of the \textit{rms} surface roughness $\rho$ to the indentation depth becomes small. The effect of the MCI tangential compliance has been recently probed experimentally \cite{Berthoud-Baumberger-ProcRSocA-1998, Bureau-Caroli-Baumberger-ProcRSocLondA-2003} : a global reversible deformation of the interface between two contacting solids is measured before irreversible slippage occurs ; regardless of the normal load, the maximum shear deformation of the interface before slippage is of the order of its \textit{rms} roughness $\rho$.

In the present set of experiments, these two effects can be uncoupled since the tangential stress at the interface vanishes under lubricated conditions. The fact that the lubricated profiles ($`[!]$ on fig.\ref{Pmax-EPL1}) are compatible with those calculated for a frictionless contact suggests that the MCI normal compressibility effect is negligible in most of the load range explored (typically for $P \gtrsim 1 \,N$). Still, it is probably responsible for the fact that $S_{max} / p^0_{max}$ falls below $1$ for the smallest loads when the indentation depth becomes comparable to the thickness $\rho$ of the rough layer.

The tangential compliance of the interface is thus expected to be primarily responsible for the observed drop from $\mu_{macro}$ to $\mu_e$ for dry contacts. This effect can be qualitatively understood by first considering the case of an interface with infinite friction. The normal loading of a sphere-on-plane contact yields a divergence of the shear stress at the edge of the contact region \cite{Spence-JElasticity-1975, Johnson-CUP-1985}. In a system with finite friction, this stress is relaxed by the development of a slip annulus at the periphery of the contact region which coexists with a central (circular) stick region, as discussed by Spence \cite{Spence-JElasticity-1975} and observed in our FEM calculations. The radius of the stick region is a growing function of the friction coefficient. As compared to a rigid interface, the existence of a finite tangential compliance allows for a subsequent strain relaxation at the elastomeric surface which is qualitatively equivalent to reducing the value of the apparent friction coefficient. This effect should vanish as the ratio of the tangential displacement induced by the sphere-on-plane loading becomes larger than the maximum elastic displacement $\rho$ allowed by the MCI deformation. The effective friction coefficient is thus expected to asymptotically reach $\mu_{macro}$ as the load $P$ is increased in qualitative agreement with our observations (fig.\ref{Pmax-EPL1}). It is noticeable however that the reduction of the apparent friction coefficient remains significant even for loads associated with a normal displacement of the rigid substrate more than $20$ times larger than the interface thickness $\rho$.

The results reported in this Letter provide evidence for the inadequacy of local Amontons' friction law to correctly capture sphere-on-plane MCI pressure profiles. They provide the first quantitative experimental measurements of such stress fields over a large range of loads for both lubricated and frictional contacts. It appears that each profile can be correctly described within the scope of Amontons' rigid-plastic law, but with a load-dependent effective friction coefficient. This coefficient grows with the total applied load but remains much smaller than the macroscopic friction coefficient even when the standard smooth hypothesis criterion (\textit{rms} surface roughness much smaller than the indentation depth) is reached. By comparing lubricated and dry frictional contacts, we can separate the effects of normal and tangential compliance of the MCI and conclude that the latter is primarily responsible for the observed deviation to the local Amontons' law.

More generally, this study provides experimental support to previous works aiming at taking into account the rheological behaviour of the micro-contacts population in the mechanical studies of multicontact interfaces under tangential load \cite{Tworzydlo-Cecot-Oden-Yew-Wear-1998, Sellgren-Olofsson-ComputMethodsApplMechEngrg-1999, Cochard-Bureau-Baumberger-JApplMech-2003}. From the present results we suggest that it should also be done for contacts under purely normal load. A first step could be the use of an elasto-plastic-like friction law instead of the classical rigid-plastic-like Amontons's one. These results were obtained with an original MEMS-based stress sensor, which has proven to be well-suited for the study of stress fields in centimeter-sized contacts and could be used to test directly any other mechanical model of the frictional interface. Many other aspects of contact mechanics are likely to be probed with the same method, such as the dynamical frictional regimes or the history-dependence of a contact submitted to an oscillatory tangential load below the sliding threshold. Other domains such as rheology or adhesion, where accurate spatially resolved stress measurements at interfaces are needed, could also benefit from the approach described here.

\acknowledgments
The authors thank A. Ponton and B. Ladoux for their help in the measurement of the PDMS Young's modulus and A. Chateauminois and C. Fretigny for fruitful discussions.

\end{document}